\documentclass[preprint,aps]{revtex4}

\usepackage{graphicx}
\usepackage{dcolumn}
\usepackage{bm}
\usepackage{amsmath}

\begin{document}
\title{Temporal coupled-mode theory for Fano resonance in light scattering by a
single obstacle}
\author{Zhichao Ruan and Shanhui Fan}

\affiliation{Ginzton Laboratory, Department of Electrical
Engineering, Stanford University, Stanford, California 94305}

\begin{abstract}
We present a theory for Fano interference in light scattering by
individual obstacle, based on a temporal coupled-mode formalism.
This theory is applicable for obstacles that are much smaller than
the incident wavelength, or for systems with two-dimensional
cylindrical or three-dimensional spherical symmetry. We show that
for each angle momentum channel, the Fano interference effect can be
modeled by a simple temporal coupled-mode equation, which provides a
line shape formula for scattering  and absorption cross-section. We
validate the analysis with numerical simulations. As an application
of the theory, we design a structure that exhibits strong absorption
and weak scattering properties at the same frequency.

\end{abstract}



\maketitle

\section{Introduction}

Understanding light scattering by individual particles is of
fundamental importance \cite{bohren1983absorption}. In recent year,
Fano interference effect \cite{fano1961effects} in such light
scattering has attracted significant attention. Individual particles
can support resonances. In the vicinity of such resonance, the
spectrum of either scattering cross-section, or  extinction
cross-section, can exhibit Fano line shape, where the spectrum
varies asymmetrically with respect to resonant frequency
\cite{tribelsky2008light,fedotov2007sharp,luk2007peculiarities,christ2008symmetry,liu2008magnetoinductive,hao2008symmetry}.
Such Fano interference effects are also closely related to the
classical analogue to electromagnetically induced transparency (EIT)
\cite{harris1997EIT}, which have recently been proposed and
demonstrated in scattering systems consisting of metallic elements
\cite{zhang2008plasmon,tassin2009low,yannopapas2009electromagnetically,tassin2009planar,papasimakis2008metamaterial,papasimakis2009metamaterial,liu2009plasmonic}.
Although the Fano effect in particle scattering has been
demonstrated in different systems, given the ubiquitous nature of
such effect, it would be very useful to present a broader theory
that captures some of general aspects.

In this paper, we generalize the temporal coupled-mode theory---
previously developed to account for Fano interference effect in
waveguide or grating systems \cite{fan2003temporal}--- to treat
light scattering by individual resonant obstacle. In the theory of
\cite{fan2003temporal}, the Fano line shape appears in the
transmission spectrum of the system. For individual obstacle,
transmission spectrum is no longer well defined. Instead, the theory
here aims to calculate the scattering cross-section of the obstacle.
Also related to this work, the temporal coupled mode theory has
recently been generalized to treat small particle scattering in
Ref.~\cite{hamam2007coupled}. The work of
Ref.~\cite{hamam2007coupled} however, did not include Fano
interference effect.

The paper is organized as follows: In Section 2, we show that for
each angle momentum channel, the Fano interference effect can be
modeled by a simple temporal coupled-mode equation, which provides a
line shape formula for scattering  and absorption cross-section. In
Section 3, we compare the theoretical predictions to numerical
simulations. Finally, in Section 4, as an application of the theory,
we design a structure that exhibits strong absorption and weak
scattering properties at the same frequency.

\section{Theory}
\subsection{Brief review of standard scattering theory}
We start by briefly reviewing the standard scattering theory for a
single obstacle. For simplicity, this paper only discusses  the
two-dimensional (2D) case where the obstacle is uniform in the $z$
direction, but the concept is straightforwardly generalizable to the
three-dimensional cases. Consider an obstacle located at the origin,
surrounding by air. When a TM wave (with its magnetic field $H$
polarized along the $z$-direction) impinges on the obstacle, the
total field in the air region outside the scatterer can be written
as:
\begin{equation}
   {{H_{total}}}= {\sum\limits_{l =  - \infty }^\infty  {H_0 \left( {h_l^ + H_l^{(2)}({k}\rho )\exp (il\theta ) + h_l^ - H_l^{(1)}({k}\rho )\exp (il\theta )} \right)} }
  \label{eq:total}
\end{equation}
where $H_0$ is a normalization constant, $(\rho,\theta)$ is the
polar coordinates oriented at the origin, $k$ is the wave number in
air, and $H_l^{(1)}$ ($H_l^{(2)}$) is the $l$-th order Hankel
function of the first (second) kind. Since $H_l^{(1,2)}(k\rho ) \to
\sqrt {\frac{2}{{\pi k}}} {e^{i(\frac{{l\pi }}{2} + \frac{\pi
}{4})}}\frac{{{e^{( + , - )ik\rho }}}}{{\sqrt \rho  }}$  when $\rho
\to \infty $, and taking the convention that the field varies in
time as $\exp(-i \omega t)$, one can identify $h_l^+$ and $h_l^-$ as
the incoming and outgoing wave amplitudes. The power carried in such
incoming (or outgoing) wave is
\begin{equation}
P_l^ \pm  = \frac{2 }{{\omega {\varepsilon _0}}}{\left| {{H_0}}
\right|^2}{\left| {h_l^ \pm } \right|^2}
\end{equation}
(Notice that the power in two-dimension has the unit of $\rm{W/m}$.)
By choosing the magnetic field normalization
\begin{equation} {H_0}
= 1\sqrt {\rm{{\frac{W}{m}}}\frac{{\omega {\varepsilon _0}}}{2 }}
\end{equation}
we have then $\left| {h_l^ + } \right|^2$ and $\left| {h_l^ - }
\right|^2$ representing the incoming and outgoing power carried by
waves in the $l$-th angular momentum channel, measured in the unit
of $\rm{W/m}$.

In the cases where the scatterer is much smaller than the incident
wavelength, or the system has cylindrical symmetry, the $l$-th order
incoming wave only excites the same order outgoing wave. So we
define $R_l$ as
\begin{equation}
\label{eq:Rl} {R_l} \equiv \frac{{h_l^ - }}{{h_l^ + }},
\end{equation}
which can be thought of as a ``reflection coefficient" since it
relates the outgoing wave to the incoming wave in each channel.
Moreover, if the obstacle is lossless, the power carried by the
outgoing wave must be equal to that of the incoming wave.
Consequently,
\begin{equation} \label{eq:Rl} {R_l}  = {e^{i\phi_l }}
\end{equation}
where $\phi_l$ is a real phase factor.

We will now calculate the scattering, absorption, and extinction
cross-sections of an obstacle. For this purpose, imagining a plane
wave incident upon the obstacle, the scattering (absorption)
cross-section is then defined as the total power scattered
(absorbed), divided by the intensity of the incident field. For 2D
system the cross-section has the unit of length. Mathematically, for
this scenario, we write the total field in the air region outside
the scatterer as
\begin{equation}
\label{eq:total2} {H_{total}} = {H_0}\left( {\exp (i{\bf{k}} \cdot
{\bf{r}}) + \sum\limits_{l =  - \infty }^\infty
{{i^l}{S_l}H_l^{(1)}(k\rho )\exp (il\theta )} } \right) ,
\end{equation} where
${\bf{k}}$ is the wave vector of the incident plane wave, $S_l$ is
related to the scattered field and referred to as the scattering
coefficient. To connect to Eq.~(\ref{eq:total}), the plane wave is
expanded as
\begin{equation}
\label{eq:total3} \exp (i{\bf{k}} \cdot {\bf{r}}) = \sum\limits_{l =
- \infty }^\infty  {{i^l}\frac{1}{2}\left( {H_l^{(1)}(k\rho ) +
H_l^{(2)}(k\rho )} \right) \exp (il\theta )}. \end{equation}
Combining Eqs.~(\ref{eq:total2}) and (\ref{eq:total3}), and
comparing to Eq.~(\ref{eq:total}), we have
\begin{equation}
\label{eq:relation} {S_l} = \frac{{{R_l} - 1}}{2}.
\end{equation}
Furthermore, in the $l$-th channel, the scattered power $P_{sct}$
and the absorbed power $P_{abs}$ are
\begin{equation}
\begin{array}{l}
 {P_{sct}} = \frac{2}{{\omega {\varepsilon _0}}}{\left| {{S_l}} \right|^2}\left| {H_0} \right|^2 \\
 {P_{abs}} = \frac{1}{{\omega {\varepsilon _0}}}\left( {1 - {{\left| {{R_l}} \right|}^2}} \right)\left| {H_0} \right|^2 = \frac{2}{{\omega {\varepsilon _0}}}\left( { - {\mathop{\rm Re}\nolimits} \{ {S_l}\}  - {{\left| {{S_l}} \right|}^2}} \right)\left| {H_0} \right| ^2\\
 \end{array}
\end{equation}
Following the definition of the scattering cross-section, the
contribution to the total scattering cross-section from the $l$-th
channel is
\begin{equation}
C_{sct,l} \equiv \frac{{{P_{sct}}}}{{{I_0}}} = \frac{{\frac{2
}{{\omega {\varepsilon _0}}}{{\left| {{S_l}} \right|}^2}{{\left|
{{H_0}} \right|}^2}}}{{\frac{1}{2}\sqrt {\frac{{{\mu
_0}}}{{{\varepsilon _0}}}} {{\left| {{H_0}} \right|}^2}}} =
\frac{{2\lambda }}{ \pi}{\left| {{S_l}} \right|^2},\end{equation}
and the total scattering cross-section is
\begin{equation}
\label{eq:cross-section}
 {C_{sct}} = \frac{{2\lambda }}{\pi }\sum\limits_{l =  - \infty }^\infty  {{{\left| {{S_l}} \right|}^2}},  \\
\end{equation}
where $\lambda$ is the wavelength in air. In the same way, we have
the total absorption cross-section as
\begin{equation}
\label{eq:abs-cross-section} {C_{abs}} = - \frac{{2\lambda }}{\pi
}\sum\limits_{l =  - \infty }^\infty  ({  {\mathop{\rm Re}\nolimits}
\{ {S_l}\} }  + {\left| {{S_l}} \right|^2})\end{equation} Finally,
the total extinction cross-section, as the sum of the the scattering
cross-section and the  absorption cross-section,  is
\begin{equation}
\label{eq:cross-section2}
 {C_{ext}} =  - \frac{{2\lambda }}{\pi }\sum\limits_{l =  - \infty }^\infty  {{\mathop{\rm Re}\nolimits} \{ {S_l}\} }
\end{equation}

\subsection{Coupled mode theory for light scattering}

The goal of our theory is then to provide a general view about these
cross-sections. In the following we consider the case that the
obstacle supports resonance for the $l$-th channel. In this
circumstance, Fano effect is the result of interference of two
pathways: the direct ``reflectance'' of the incoming wave that forms
the background, and the outgoing radiation from the excited
resonance. Using  the temporal coupled-mode theory formalism
\cite{fan2003temporal,hamam2007coupled,haus1984waves}, the dynamic
equation for the amplitude $c$ of the resonance is
\begin{equation}
\label{eq:coupled}
\begin{array}{*{20}{l}}
 \frac{{dc}}{{dt}} & = & \left( { - i{\omega _0} - \gamma_0 - \gamma } \right)c + \kappa h^ +  \\
 {h}^ -  & = & {B}h^+  + \eta c \\
 \end{array}
\end{equation}
where $\omega_0$ is the resonant frequency, $\gamma_0$ is the
intrinsic loss rate due to for example material absorption, $\gamma$
is the external leakage rate due to the coupling of the resonance to
the outgoing wave, ${B}$ is the background reflection coefficient,
and $\kappa$ and $\eta$ correspond to the coupling constant between
the resonance, and the incoming or outgoing wave, respectively. In
this section, as we consider only the $l$-th channel, for notation
simplicity we suppress the subscript $l$ in all variables in
Eq.~(\ref{eq:coupled}). Here the amplitude $c$ is normalized such
that ${\left| c \right|^2}$ corresponds to the energy inside the
resonator \cite{haus1984waves}. Note that such coupled-mode
formalism is, strictly speaking, valid only when ${\gamma _0} +
\gamma \ll {\omega _0}$ \cite{haus1984waves}.


The coupling constants $\kappa$ and $\eta$ are related to each other
by energy conservation and time-reversal symmetry considerations
\cite{fan2003temporal}. In the lossless case (i.e. $\gamma_0=0$), if
the incoming wave is absent (i.e. $h^+=0$),  from
Eq.~(\ref{eq:coupled}) we have
\begin{equation}
\label{eq:decay}
\begin{array}{l}
 c = A\exp ( - i{\omega _0}t - \gamma t) \\
 {h}^ -  = A\eta \exp ( - i{\omega _0}t - \gamma t) \\
 \end{array}\end{equation}
where $A$ is an arbitrary constant. According to energy
conservation, the energy leakage rate must be equal to the power of
the outgoing wave, i.e.
\begin{equation}
\frac{{{\rm{d}}{{\left| c \right|}^2}}}{{{\rm{d}}t}} =  - 2\gamma
{{\left| c \right|}^2} =  - {\left| {h^ - } \right|^2} =  - \eta
{\eta ^*}{{\left| c \right|}^2}
\end{equation}
which requires that
\begin{equation}
\label{eq:nu}
 \eta {\eta ^*} = 2\gamma
\end{equation}

Now let us perform a time-reversal transformation for the
exponential decay process as described by Eq.~(\ref{eq:decay}). The
time-reversed case is represented by feeding the resonator with
exponentially growing wave amplitude $(h^-(-t))^*$. Such excitation
results in a resonant amplitude  $(c(-t))^*$ that also grows
exponentially, without the outgoing wave. Described this
time-reversed scenario using Eq.~(\ref{eq:coupled}), we have
\begin{subequations}
\begin{equation}
\label{eq:coeff4}
 \kappa {\eta ^*} = 2\gamma
\end{equation}
\begin{equation}
 B{\eta ^*} + \eta  = 0
\end{equation}
\end{subequations} Comparing Eq.~(\ref{eq:coeff4}) to (\ref{eq:nu}), we obtain
\begin{equation}
\label{eq:coeff5}
 \kappa = \eta
\end{equation}

Eqs.~(\ref{eq:nu})-(\ref{eq:coeff5}) are the main results of the
Fano interference theory. If the system is lossless, the background
reflection is $B = e^{i \phi}$ and therefore we can determine
$\kappa$ and $\eta$ as
\begin{equation}
\label{eq:coeff2} \kappa  = \eta  = \sqrt {2\gamma }
{e^{i(\frac{\phi }{2} + \frac{\pi }{2} - n\pi )}}
\end{equation}
where $n$ is an arbitrary integer.  For lossy systems, one should
expect that the strongest contribution of the loss is to the
resonant properties. In such a case, as an ansatz, we will introduce
a non-zero intrinsic loss rate $\gamma_0$ in Eq.~(\ref{eq:coupled}),
while still approximating the background scattering as a lossless
process. We will see below in the numerical example that such an
approach is in fact sufficient for scattering from lossy plasmonic
particles. A more rigorous approach, which we will not pursue here,
will be to treat explicitly the loss as coupling to an additional
decay channel, the combined system with the additional decay channel
can then be treated using theory that is similar to
Eq.~(\ref{eq:coupled}) but without  the need of introducing the
intrinsic loss rate.

We now suppose that the incident wave has a frequency $\omega$. The
reflection coefficient $R$ can be straightforwardly obtained using
Eqs.~(\ref{eq:coupled}) and (\ref{eq:coeff2}):
\begin{equation}
\label{eq:reflection-coef}
\begin{array}{*{20}{l}}
 R & = & \frac{{h^ - }}{{h^ + }} = {e^{i\phi }} + \frac{{ - 2\gamma {e^{i\phi }}}}{{i({\omega _0} - \omega ) + \gamma_0 + \gamma }} \\
{}  & = & \frac{i({\omega _0} - \omega )+ \gamma_0  - \gamma }{{i({\omega _0} - \omega ) + \gamma_0 + \gamma }} {e^{i\phi }}\\
 \end{array}
\end{equation}
One can easily identify that in the lossless case (i.e.
$\gamma_0=0$), the amplitude of $R$ is unity  as expected. Following
Eq.~(\ref{eq:relation}), the scattering coefficient is
\begin{equation}
\label{eq:coeff3}
\begin{array}{*{20}{l}}
 {S} & = & \frac{1}{2}({R} - 1) \\
{}  & = & \frac{1}{2}\frac{{(i({\omega _0} - \omega )+ \gamma_0)({e^{i\phi }} - 1) - \gamma (1 + {e^{i\phi }})}}{{i({\omega _0} - \omega ) + \gamma_0 + \gamma }} \\
 \end{array}
\end{equation} Substituting
Eqs.~(\ref{eq:coeff3}) into (\ref{eq:cross-section}) and
(\ref{eq:abs-cross-section}), we have the scattering and the
absorption cross-sections as
\begin{subequations}
\label{eq:coss-section-form}
\begin{equation}
\label{eq:coeff6} {C_{sct}} = \frac{{2\lambda }}{\pi }{\left|
{\frac{1}{2}\frac{{(i({\omega _0} - \omega ) + {\gamma
_0})({e^{i\phi }} - 1) - \gamma (1 + {e^{i\phi }})}}{{i({\omega _0}
- \omega ) + {\gamma _0} + \gamma }}} \right|^2}
\end{equation}
\begin{equation}
\label{abs-coss-section-form} {C_{abs}} = \frac{{2\lambda }}{\pi
}\frac{{{\gamma _0}\gamma }}{{{{(\omega  - {\omega _0})}^2} +
{{({\gamma _0} + \gamma )}^2}}}
\end{equation}
\end{subequations}

\subsection{General line shapes in a single channel}

\begin{figure}[h]
\centerline{\includegraphics[width=3.8in]{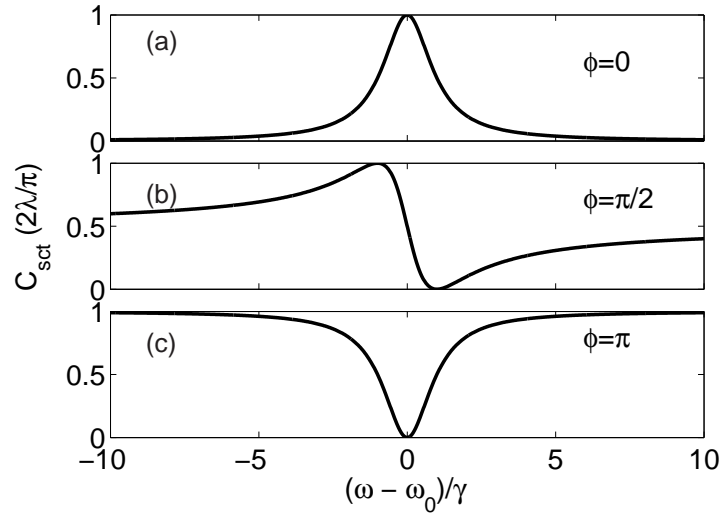}}
 \caption{\label{fig:fano} Scattering cross-section spectrum as given by
Eq.~(\ref{eq:coeff3}) for the lossless case ($\gamma_0 =0$). (a-c)
correspond to $\phi=0, \pi/2, \pi$, respectively.}
\end{figure}

We now use Eq.~(\ref{eq:coss-section-form}) to illustrate general
line shapes of scattering and absorption  cross-section spectra in a
single channel. Let us first discuss the scattering cross-section.
In the lossless case where $\gamma_0=0$, Fig.~\ref{fig:fano} plots
the scattering cross-section spectra with different phase factor
$\phi$ as arised from the background ``reflection".

\begin{figure}[h]
\centerline{\includegraphics[width=3.8in]{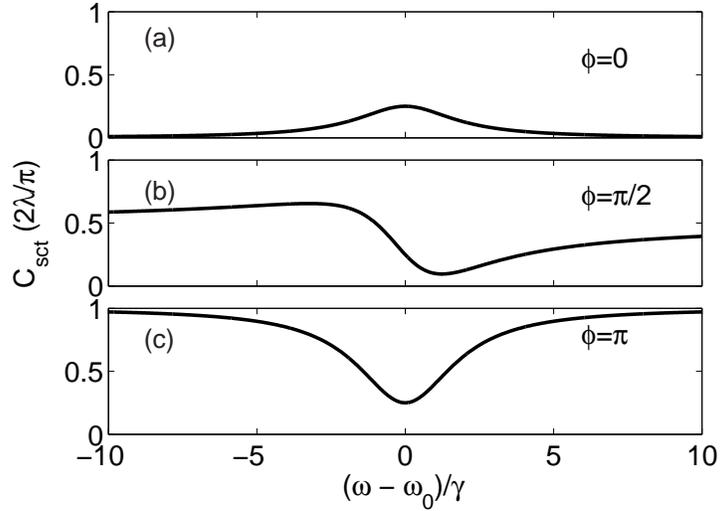}}
\caption{\label{fig:fanolossy}  Scattering cross-section spectrum as
given by Eq.~(\ref{eq:coeff3}) for the lossy case with
$\gamma_0=\gamma$.}
\end{figure}

For the phase factor $\phi=0$, the spectrum is a Lorentizian
function [Fig.~\ref{fig:fano}(a)]. Note that the resonance creates
maximal scattering at the resonant frequency $\omega_0$. Since $\phi
= 0$ corresponds to a very weak background scattering, in this case
the only contribution to the scattering process is from the
resonance. An example of a very weak background scattering is that
for the $H$-polarization as we consider here, when the radius of a
metallic cylinder is far smaller than the wavelength. In this case,
the presence of a surface-plasmon resonance would create a
Lorenztian line shape in the scattering cross-section spectrum, and
dramatically increase the scattering cross-section.

For all cases where $\phi$ is not equal to $0$ or $\pi$, the
spectrum is not a Lorentizian, but rather exhibits a Fano asymmetric
line shape, which exhibits both enhancement and suppression of
scattering coefficients in the vicinity of the resonance. For
example, Fig.~\ref{fig:fano}(b) shows the scattering cross-section
spectrum for $\phi=\pi/2$. We note that in the Fano line shape,
neither the minimum nor the maximum in scattering cross-section
occurs at the resonant frequency $\omega_0$. Through
Eq.~(\ref{eq:coeff6}), we can determine that the minimum and maximum
occur at the frequency $\omega_0\pm\gamma\tan(\phi/2)$ and
$\omega_0\mp\gamma\cot(\phi/2)$ respectively, where the $\pm$ sign
corresponds to $0 <\phi <\pi$ or $ -\pi< \phi < 0$. Therefore, in
order to switch the scattering cross-section from the minimum to the
maximum, the smallest frequency required is $2\gamma$ which is for
the case $\phi=\pm\pi/2$. This result is in fact consistent with a
previous study on optical switching using Fano resonances
\cite{fan2002sharp}.

When $\phi = \pi$, the background exhibits maximum scattering, and
the presence of the resonance creates a  dip at $\omega_0$
--- no scattering occurs at the resonance frequency
[Fig.~\ref{fig:fano}(c)]. This effect is closely related to the
all-optical analogue of EIT, which occurs when a super-radiant state
has the same resonant frequency as a sub-radiant state in the same
scattering channel. The super-radiant state by itself establishes a
broad scattering peak and thus can be regarded as establishing a
background in the vicinity of the sub-radiant resonance. The
presence of the sub-radiant state introduces a dip and renders the
system transparent (i.e. with zero scattering) at the resonant
frequency.  A better treatment of EIT effect however requires one to
treat the super-radiant and the sub-radiant state at an equal
footing \cite{Suh2004temporal} and will be done in later works.


To show the effect of the intrinsic loss, Fig.~\ref{fig:fanolossy}
shows the scattering cross-section spectra when $\gamma_0=\gamma$.
Comparing it to Fig.~\ref{fig:fano}, one can see that the line
shapes are similar to the corresponding lossless cases, but the
presence of the loss reduces the cross-section variation as a
function of frequency. In particular, the scattering cross section
can no longer reach $2 \lambda / \pi$, which is the maximal
contributions possible in a single channel. Neither can the
scattering cross-section reaches zero. In the presence of loss it is
no longer possible to completely eliminate scattering effect at a
single frequency through the use of a Fano interference with a
single resonance.

We now consider the absorption cross-section spectrum described by
Eq.~(\ref{abs-coss-section-form}). In contrast to the scattering
cross-section spectrum, which varies with the background phase
factor, the absorption cross-section is independent of the phase
factor. Instead, it always has a symmetric Lorentzian line shape
with its maximum at the resonant frequency.  Thus, Fano interference
effect does not affect the absorption properties of  the obstacle.

\section{Numerical Validation}

\begin{figure}[h]
\centerline{\includegraphics[width=2.8in]{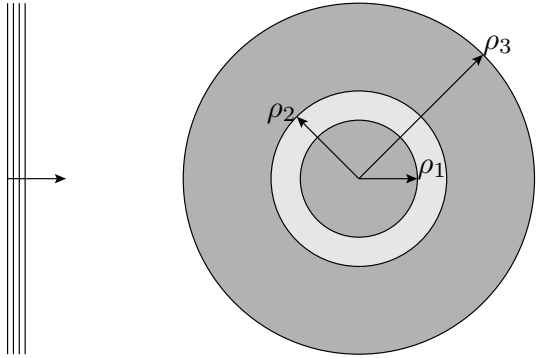}}
\caption{\label{fig:cavity} Schematic of a metal-dielectric-metal
cylinder obstacle. The dark gray area is a metal described by a
Drude model. The light gray area is dielectric ($\epsilon_d = 12.96
$). The surrounding media is assumed as air.}
\end{figure}

\begin{figure}[h]
\centerline{\includegraphics[width=5.8in]{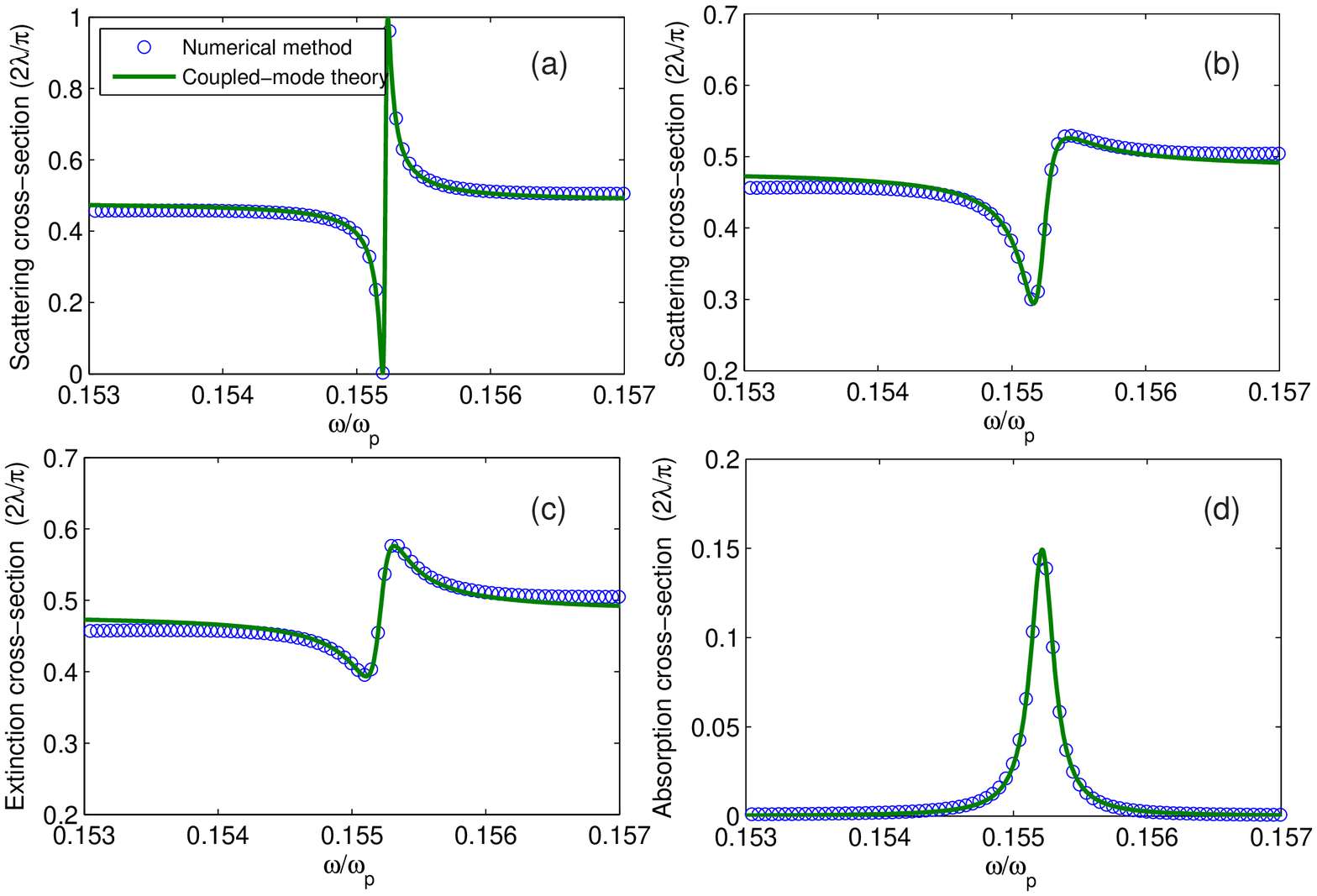}}
\caption{\label{fig:single} Cross-section of the $l=0$ channel for
the scatterer shown in Fig.~\ref{fig:cavity}, where the geometric
parameters are $\rho_1=0.285\lambda_p$, $\rho_2=\lambda_p$, and
$\rho_3=1.5\lambda_p$ ($\lambda_p$ is the wavelength  in the vacuum
at the plasma frequency $\omega_p$). (a) Scattering cross-section
for the lossless metal case of $\gamma_d=0$; (b-d) the  scattering,
extinction, and absorption cross-section spectra for the lossy metal
case of $\gamma_d=0.001\omega_p$.}
\end{figure}

To check the validity of the above theory, we compare
Eq.~(\ref{eq:coss-section-form}) to the results of numerical
simulations based on the Lorentz-Mie method of a cylindrical
scatterer. Figure~\ref{fig:cavity} shows the schematic of the
scatterer, which consists of multiple concentric  layers of metal
and dielectric. The permittivity of dielectric is
$\varepsilon_{d}=12.96$, and the metal is described by a Drude model
${\varepsilon _m} = 1 - {{\omega _p^2}}/({{{\omega ^2} + i{\gamma
_d}\omega }})$ ($\omega_p$ and $\gamma_d$ are the plasma frequency
and the damping rate, respectively). In the frequency range of
$\omega < \omega_p$, a corresponding planar structure can support
surface-plasmon waveguide modes that are confined in the dielectric
region. For the cylindrical structure, whispering gallery modes
related to such waveguide modes can be formed in the low angular
momentum channels \cite{Catrysse2009Simple}.

We start with the lossless case where $\gamma_d=0$. For the geometry
parameters of $\rho_1=0.285\lambda_p$, $\rho_2=\lambda_p$, and
$\rho_3=1.5\lambda_p$ (where $\lambda_p= 2 \pi c/ \omega_p$, with
$c$ being the speed of light in vacuum), the scattering
cross-section for the $l=0$ channel is plotted as circles in
Fig.~\ref{fig:single}(a). It shows a typical Fano resonant line
shape around $0.155\omega_p$.

To compare to the analytic results, we determine the parameters
required in Eq.~(\ref{eq:coss-section-form}) in the following way:
We express the $H$ field in each layer as a linear superposition of
$J_0(k\rho)$ and $H^{(1)}_0(k\rho)$, except in the innermost layer
where the field is proportional to only $J_0$, and outermost layer
where the field is proportional to only $H^{(1)}_0$. By matching the
boundary condition at all metal-dielectric interfaces, we obtain a
transcendental equation with frequency as a variable. By solving for
the complex roots of the transcendental equation, we determine the
resonance frequency (the real part of the root) and the leakage rate
(the imaginary part of the root). For the structure in shown here,
in the lossless case, we obtain $\omega_0 = 0.1552\omega_p$, $\gamma
= 1.9166\times10^{-5}\omega_p$. The phase factor $\phi= -0.4882\pi$
is established by calculating the scattering coefficient of a
uniform metallic cylinder with the same size. Using these
parameters, the scattering cross-section calculated by
Eq.~(\ref{eq:coeff6}) shows an excellent agreement with the
numerical result [Fig. ~\ref{fig:single}(a)] .

We now introduce a metal loss characterized by
$\gamma_d=0.001\omega_p$. Repeating the same process as outlined
above for the lossless case, we determine a resonant frequency
$\omega_0 = 0.1552\omega_p$ and a total loss rate $\gamma + \gamma_0
= 1.0492\times10^{-4}\omega_p$. Assuming that the leakage rate is
the same as the lossless case, i.e.
$\gamma=1.9166\times10^{-5}\omega_p$, we have the intrinsic loss
rate $\gamma_0=8.5749\times10^{-5}\omega_p$. The background phase
shift $\phi$ is determined using the same procedure by considering
the scattering of a corresponding uniform metallic cylinder, which
results in $\phi=-0.4882\pi+8.6\times10^{-4}i$. Comparing with the
lossless case, we notice that the background phase shift is not
significantly affected by introducing loss in the system.  Thus, in
the theory for the lossy case, we will choose as a parameter the
background phase shift $\phi = - 0.4882 \pi$, and ignore the small
loss that occurs in the background. Using these parameters, the
theoretical spectra again agree well with the numerical spectra
[Figs.~\ref{fig:single}(b-d)].  The example here thus provides a
validation of the theory in the single channel case.

\section{Multiple-channel resonant obstacle}

\begin{figure}[h]
\centerline{\includegraphics[width=3.4in]{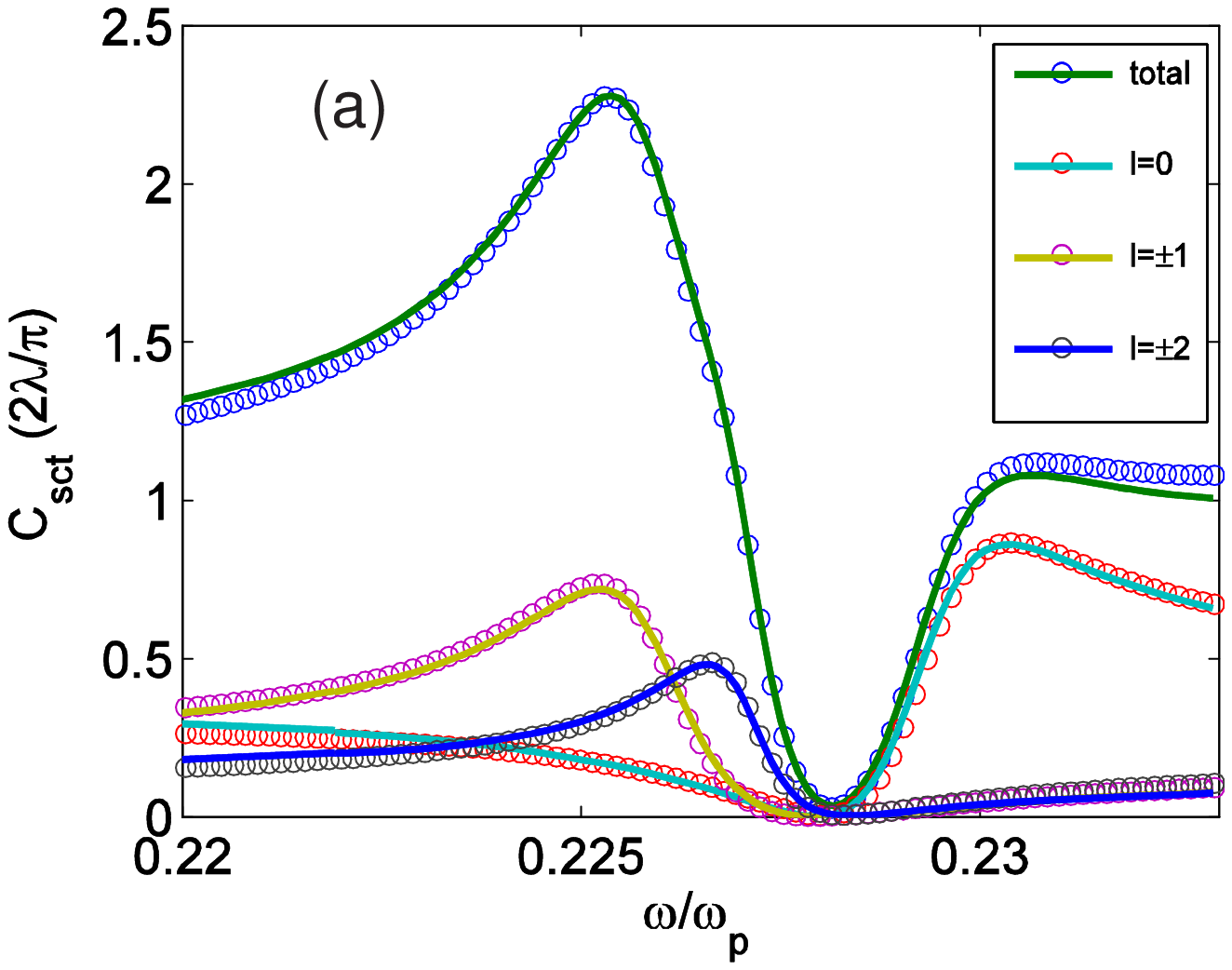}}
\centerline{\includegraphics[width=3.4in]{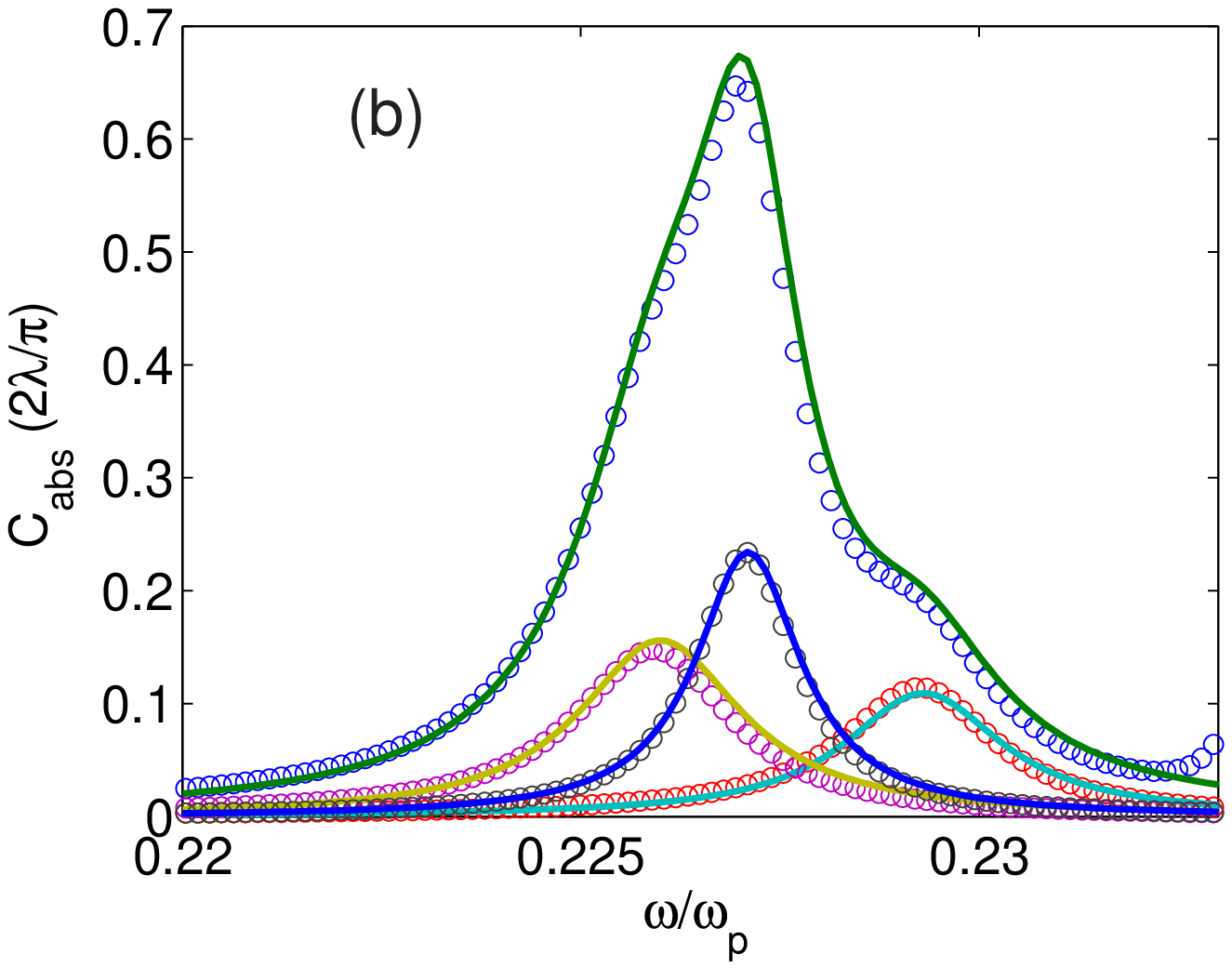}}
\caption{\label{fig:multiple} Scattering (a) and absorption (b)
cross-section spectra for the scatterer shown in
Fig.~\ref{fig:cavity}, with the geometric parameters of
$\rho_1=0.36\lambda_p$, $\rho_2=0.73\lambda_p$ and
$\rho_3=\lambda_p$, and the metal damping rate of
$\gamma_d=0.001\omega_p$. The circles are from the Lorentz-Mie
method, and the solid lines are the theoretical fit using
Eq.~(\ref{eq:coeff3}).}
\end{figure}

The theory developed in Section II can be straightforwardly
generalized to include resonances in multiple channels
(where the resonance in each channel
can only be excited by the same channel incoming wave
and coupled to the same channel outgoing wave), since the
total cross-section is the sum of the contributions from all
channels. As an example, here we apply our theory to  a
metal-dielectric-metal cylinder scatterer where all channels that
strongly contribute to the scattering process exhibit the Fano
effect.

We again consider the structure is shown in Fig.~\ref{fig:cavity},
with a new set of  parameters $\rho_1=0.36\lambda_p$,
$\rho_2=0.73\lambda_p$ and $\rho_3=\lambda_p$. The metal has a
damping rate $\gamma_d=0.001\omega_p$. The circles in
Figs.~\ref{fig:multiple}(a), and (b) correspond to the scattering
and absorption cross-section spectra, respectively, as obtained from
numerical simulations. In the frequency range of $0.22\sim0.233
\omega_p$, the total cross-section of the system is dominated by the
contributions from five channels of $l=0,\pm1,\pm2$. For each
channel, the scattering cross-section spectrum and the absorption
cross-section spectrum show the Fano and Lorentizian line shape,
respectively. We use the above theory to fit these curves. The
fitting results are plotted as the solid lines in
Fig.~\ref{fig:multiple}, which again indicates good agreement
between theory and simulation.

We note that this obstacle can be potentially used as a cloaked
sensor \cite{alu2009cloaking}--- when the obstacle is placed in an
electromagnetic field, it absorbs the energy but creating only
minimum scattering within a narrow range of frequencies. From
Fig.~\ref{fig:multiple}, one can see that at the frequency
$\omega=0.2282\omega_p$, the total scattering cross-section of the
obstacle is only $0.03(2\lambda/\pi)$, while the total absorption
cross-section is $0.32(2\lambda/\pi)$.  In the presence of loss, the
scattering of the particle cannot be completely eliminated in this
case, as can be seen from Eq.~(\ref{eq:reflection-coef}).
Nevertheless, the scattering can be substantially reduced as we show
here. The Fano interference effect thus provides an interesting
alternative mechanism for creating a cloaked sensor as compared to
Ref.~\cite{alu2009cloaking}.

\section{Summary and outlook}

In summary, we present a theory for Fano interference in light
scattering by individual obstacle, based on a temporal coupled-mode
formalism. We show that for each angle momentum channel, the Fano
interference effect can be modeled by a simple temporal coupled-mode
equation, which provides a line shape formula for scattering  and
absorption cross-section. We validate the analysis with numerical
simulations. As an application, we design a structure that exhibits
strong absorption and at the same time weak scattering properties.

We note that this theory is applicable for obstacles that are much
smaller than the incident wavelength, or systems that have
two-dimensional cylindrical or  three-dimensional spherical
symmetry. For arbitrary-shaped obstacles or assembled clusters, the
overall scattering matrix of the system is no longer diagonalizable
in the angular momentum basis. Instead, for a theory of Fano
interference, one may define the channel as the eigenstates of the
background scattering matrix, the resonance in this case is likely
then to couple to several scattering channels simultaneously. Such a
theory will be developed in future research.

This work is supported by DARPA/MARCO under the Interconnect Focus
Center, by AFOSR Grant No. FA9550-04-1-0437, and by the DOE Grant
No. DE-FG 07ER46426.


\end{document}